\documentclass[aps,nofootinbib,showpacs
]{revtex4}
\usepackage[cp1251]{inputenc}
\usepackage{amsmath}
\usepackage{amsfonts}
\usepackage{amssymb}
\usepackage{graphicx}
\usepackage{epsf}
\usepackage{wrapfig}
\usepackage[mathscr]{eucal}
\newcommand{\ud}{\mathrm{d}}

\begin{document}
\title{The holographic screen at low temperatures}
\author{V.V.Kiselev, S.A.Timofeev}
\affiliation{Russian State Research Center ``Institute for High
Energy Physics'',
Pobeda 1, Protvino, Moscow Region, 142281, Russia\\
Moscow Institute of Physics and Technology, Institutskii per. 9,
Dolgoprudnyi, Moscow Region, 141701, Russia}
\begin{abstract}
A permissible spectrum of transverse vibrations for the holographic screen
modifies both a distribution of thermal energy over bits at low temperatures
and the law of gravitation at small accelerations of free fall in
agreement with observations of flat rotation curves in spiral galaxies. This
modification relates holographic screen parameters in de Sitter space-time
with the Milgrom acceleration in MOND.
\end{abstract}
\maketitle

\section{Introduction}

The holographic approach \cite{Holograph,Holograph1,B-rev} within the thermodynamical
formulation of gravity
\cite{Bekenstein,Bekenstein1,Bekenstein2,Hawking,Hawking1,Hawking2,Hawking3,Hawking4,Jacobson,Padmanabhan,Padmanabhan1} became logically
accomplished in papers by E.Verlinde \cite{Verlinde} and T.Padmanabhan
\cite{Padmanabhan2,otherPadmanabhan} as well as in further developments,
modifications and applications of their ideas in various aspects
\cite{develope,develope1,develope2,develope3,iCosmo,iCosmo1,iCosmo2,iCosmo3,iCosmo4,iCosmo5,iCosmo6,iCosmo7,iCosmo8,iCosmo9,iCosmo10,iCosmo11,iCosmo12,iCosmo13, iNewton,iNewton1,iNewton2,iNewton3,iNewton4,iNewton5,iRindler,iRindler1,iRindler2,iRindler3,specific,specific1,specific2,iLQG,iLQG1,iLQG2,Maekela,Morozov-kin,other,other1}.

In \cite{KS-gol} we have connected the entropic force acting on a probe
particle near a holographic screen with its surface tension $p_A$. Therefore,
we have calculated the surface density of holographic entropy $s_A=\ud S/\ud
A$ as follows from the equipartition rule, so that
\begin{equation}
 \label{1-4}
      s_A=\frac{1}{4G},
\end{equation}
whereas the variation of screen energy $E$ is given by the following
relation:
\begin{equation}
 \label{dE}
      \ud E= T\ud S-p_A\ud A,
\end{equation}
with $A$ denoting the screen area, while the surface tension is equal to
\begin{equation}
 \label{pA}
      p_A=-T\, s_A.
\end{equation}

However, according to the Nernst heat theorem one should expect that the
surface entropy density tends to zero at zero temperature: $s_A\to 0$ at
$T\to 0$, because dynamical degrees of freedom are frozen out at absolute
zero. Furthermore, according to \cite{landauV} the derivative of pressure
with respect to the temperature at a constant volume should tend to zero, so
that in the case under consideration one gets
\begin{equation}
 \label{dif-pA}
      \left(\frac{\partial p_A}{\partial T}\right)_{\hskip-3pt A}\to 0,\quad
\mbox{at}\quad
      T\to 0.
\end{equation}
From (\ref{1-4}) and (\ref{pA}) it follows that
\begin{equation}
 \label{pA2}
      \left(\frac{\partial p_A}{\partial
      T}\right)_{\hskip-3pt A}=-\frac{1}{4G}=\mbox{const.}
\end{equation}
That is in conflict with the Nernst heat theorem, and the same concerns for
expression (\ref{1-4}) determining the entropy density, which is independent
of the temperature.

Obviously, one could satisfy the Nernst heat theorem if the degrees of
freedom on the holographic screen are frozen out when approaching the
absolute zero. Therefore, the equipartition distribution of energy is
modified as
\begin{equation}
 \label{mod}
      E=\frac12\,TN\quad\mapsto\quad E=\frac12\,TNB(T),
\end{equation}
where $E=M$ is the gravitating mass inside the screen, and the number of
``bits'' $N$ is equal to the number of Planck cells\footnote{The Newton
gravitation constant $G=\ell_\mathrm{Pl}^2$, where
$\ell_\mathrm{Pl}=1/m_\mathrm{Pl}$ is the Planck length inverse to the Planck
mass. Here we put $c=1$, $\hbar=1$ and $k_B=1$.} on the screen area
\begin{equation}
 \label{bits}
      N=\frac{A}{G},
\end{equation}
while
$$
      B(T)\to 0,\qquad\mbox{at}\quad T\to 0.
$$

In \cite{gol-debay} the following expression has  been heuristically
suggested
\begin{equation}
 \label{gol-debay}
      B(T)=\mathscr{D}(x),
\end{equation}
where $\mathscr{D}(x)$ is the one-dimensional Debye-function
\begin{equation}
 \label{debay}
      \mathscr{D}(x)=\frac{1}{x}\int\limits_0^x\frac{z\ud z}{\mathrm{e}^z-1},
\end{equation}
at $x=T_D/T$. The new empirical parameter $T_D$ occurs  to be comparable with
the current value of Hubble constant $H_0$. At high temperatures $T\gg T_D$
the equipartition rule is valid since $\mathscr{D}(0)=1$, but at low
temperatures $T\ll T_D$ the Debye function gets the linear behavior, namely
\begin{equation}
 \label{low}
      \mathscr{D}(x)\stackrel{x\to \infty}{\approx} \frac{T}{T_0},\qquad
      T_0=\frac{6}{\pi^2}\,T_D.
\end{equation}
It gives the modification of Newton gravitation law for a body of mass $M$,
so that in the framework of holographic formulation the acceleration of free
fall $a=2\pi T$ becomes equal to
\begin{equation}
 \label{free-accel}
      a^2=\frac{GMa_0}{r^2},
\end{equation}
where $a_0=2\pi T_0$ is the Milgrom acceleration phenomenologically
introduced in the MOND paradigm\footnote{In \cite{TeVeS} Y.Bekenstein has
constructed a relativistic Lagrangian formalism consistent with MOND.}
(modified Newtonian dynamics) \cite{Milgrom} to explain the flat rotation
curves of spiral galaxies, when the velocity of peripheral stars is
asymptotically expressed in terms of the mass of visible matter $M$ in the
galaxy according to the formula
\begin{equation}
 \label{TF}
      v_0^4=GMa_0,
\end{equation}
empirically known as the Tully-Fisher law.

Note that the authors of \cite{gol-debay} have not physically validated both
the Debye factor and the connection between the Debye temperature and Hubble
constant, though they have noted that the notion of Debye correction is
meaningful for solid states, but not for holographic screens, since the
factor describes the wide interval of temperatures from sound vibrations to
independent oscillations of lattice notes in the solid state bodies, when the
spectrum of vibrations is well approximated by the Debye formula. In
addition, the introduction of one-dimensional Debye function in
\cite{gol-debay} is the guess beyond any reasonable speculations, and it
raises a question on the dimension of holographic screen.

Here we consider collective vibrations of holographic screen
on the basis of its surface tension. We find the impossibility of propagation
of 2-dimensional sound waves with both polarization and wave vector lying in
the plane of holographic screen. So, the only permissible collective motion
of bits on the holographic screen is given by the one-dimensional scaling of
a holographic screen as whole in the transverse direction to the screen
plane\footnote{Surface oscillations of the holographic screen without
gradients of density, pressure or temperature are possible with wave vectors
tangent to the screen plane (see Appendix \ref{app1}). But because of
holographic principle, these oscillations correspond to specific vibrations
of gravitating matter inside the screen, that breaks its initial spatial
symmetry, for example, the spherical symmetry of field for a massive
point-like source.}. We analyze the frequency spectrum of such motion and
validate an appropriate thermal correction to the energy distribution law
using the Debye method, i.e. a cutoff in the frequency spectrum. We
investigate a physical interpretation for a number of vibrational modes of
the holographic screen and its connection to the scale hierarchy: the Planck
scale, the Hubble rate and a sub-Planckian scale which is of the order of the
grand unification scale in the particle physics.

\section{Collective motions of holographic screen ``bits''}

The sound speed squared $c_s^2$ in the medium is determined by the adiabatic
derivative of pressure with respect to the energy density. In the case of
holographic screen  we get
\begin{equation}
 \label{cs}
      c_s^2=\left(\frac{\partial p_A}{\partial \rho_A}\right)_{\hskip-3pt S},
\end{equation}
where the surface energy density is denoted by $\rho_A=\ud E/\ud A$ at high
temperatures, when the equipartition rule is valid, and it takes the form
\begin{equation}
 \label{rA}
      \rho_A=\frac{1}{2G}\,T=2T s_A.
\end{equation}
Taking into account the expression for the surface tension in (\ref{pA}) we
find that the speed squared for sound on the holographic screen is negative
\begin{equation}
 \label{cs2}
      c_s^2=-\frac12,
\end{equation}
whereas this expression is valid at low temperatures too because thermal
correction does not change the relation between the energy density and
surface tension $p_A=-\frac12\,\rho_A$.

Thus, the sound wave propagation along the screen with the polarization in
the screen plane is absolutely impossible. This fact can be predicted from
the definition of the holographic screen as the surface with a constant
acceleration of free fall and, hence, with a constant temperature.
Therefore, a temperature gradient is forbidden, hence, gradients of the
energy density and pressure do not propagate. The relation between the
surface tension and surface energy density confirms the validity of such the
consideration in respect to the sound waves.

The only permissible collective motion is the uniform extension or
contraction of holographic screen as a whole in the transverse direction to
the screen, i.e. the scaling transformation depending on the time: $r\mapsto
a(t) r$, where the scale factor $a(t)$ completely determines the dynamics of
such motion with single degree of freedom. In Appendix \ref{app1} we consider
the transverse oscillations of the screen propagating along the screen.
However, such oscillation with a wave vector lying in the screen plane are
unavoidably accompanied by a motion of matter inside the holographic surface.
In the case of spherically symmetric matter the vibrational wave vector
should be directed orthogonally to the screen, only.

It is easy to calculate the spectrum of such transverse motion because the
scale factor is determined by the Universe evolution. Since we are interested
of the present evolution, we can use the parametrization for the Hubble
constant $H$ in the form of
\begin{equation}
 \label{H}
      H^2(t)=H_0^2\left(\frac{\Omega_M}{a^3(t)}+\Omega_\Lambda\right),
      \qquad a(0)=1,
\end{equation}
where the energy balance in the flat Universe for fractions of the matter
$\Omega_M$ and the cosmological constant $\Omega_\Lambda$ gives
$\Omega_M+\Omega_\Lambda=1$, and numerically $\Omega_\Lambda\approx 0.76$. In
the epoch of cosmological constant the Hubble constant is equal to
\begin{equation}
 \label{HL}
      H_\Lambda = H_0\sqrt{\Omega_\Lambda},
\end{equation}
and the scale factor behaves like
\begin{equation}
 \label{scale}
      a(t)=\mathrm{e}^{H_\Lambda t},\qquad t\in (-\infty,0).
\end{equation}
The spectrum density
\begin{equation}
 \label{spectr}
      \nu(\omega)=\int\limits_{-\infty}^0\ud t\,
      \mathrm{e}^{-\mathrm{i}\omega t}a(t)=\frac{1}{H_\Lambda-\mathrm{i}\omega}
\end{equation}
gives the number of the scale-vibrational modes for the holographic screen
$N_G(\omega)$ in the frequency range $(0,\omega)$
\begin{equation}
 \label{No}
      N_G(\omega)=\int\limits_{-\omega}^{\omega}\frac{\ud
      \omega}{2\pi}\,|\nu(\omega)|=\int\limits_{0}^{\omega}\frac{\ud
      \omega}{\pi}\,\frac{1}{\sqrt{H_\Lambda^2+\omega^2}}.
\end{equation}

Notice that at small frequencies $\omega\ll H_\Lambda$ the frequency density
becomes constant, and the number of modes is equal to
\begin{equation}
 \label{No2}
      N_G(\omega)\approx \frac{\omega}{\pi H_\Lambda}\ll 1.
\end{equation}
The contribution of such vibrations to the free energy $\mathscr{F}$ can be
written as
\begin{equation}
 \label{stat}
      \mathscr{F}=-T\sum_{\mathrm{bits,}\,\omega}\ln Z(\omega),
\end{equation}
where the sum is taken over both ``bits'' on the screen and low frequency
modes of oscillations, if $N_G\ll 1$, while
\begin{equation}
 \label{Z}
      Z(\omega) =\left(1-\mathrm{e}^{-\omega/T}\right)^{-1}.
\end{equation}
At high temperatures $\omega\ll T$ the equipartition rule is reproduced
because
$$
      \ln Z\approx \ln T-\ln\omega,
$$
and
$$
      \sum_{\mathrm{bits,}\,\omega}\ln Z(\omega)\approx -\frac12 \tilde
N_\mathrm{bits}      N_G (\langle\ln \omega\rangle- \ln T)
$$
where the average value is defined as
$$
      \sum_\omega \ln \omega=N_G\langle\ln \omega\rangle,
$$
so that the energy $E=\mathscr{F}-\partial\mathscr{F}/\partial \ln T$ is
equal to
\begin{equation}
 \label{eq}
      E=\frac{1}{2}\,TN,
\end{equation}
where $N=\tilde N_\mathrm{bits} N_G$ is the total number of ``bits'' equal to
the number of Planckian cells of the screen area, and $\tilde
N_\mathrm{bits}$ is the number of ``bits'' per the scale-vibrational mode of
the screen.

At low temperatures $T\ll \omega$ by introducing the frequency cutoff
$\omega<T_D$, formula (\ref{stat}) gives
\begin{equation}
 \label{stat2}
      \mathscr{F}=\frac12\,T\tilde N_\mathrm{bits}\int\limits_0^{T_D}\frac{\ud
      \omega}{\pi H_\Lambda}\,\ln\left(1-\mathrm{e}^{-\omega/T}\right),
\end{equation}
hence, the energy equals
\begin{equation}
 \label{debay2}
      E=\frac12\,TN\,\mathscr{D}\left(\frac{T_D}{T}\right),
\end{equation}
where again $N=\tilde N_\mathrm{bits} N_G$ and $N_G=N_G(T_D)$, so that
$$
      E\approx \frac12\,N\,\frac{T^2}{T_0}, \qquad T\ll T_0.
$$
Obviously, the formulas (\ref{stat2}) and (\ref{debay2}) can be applicable
not only when $T\ll T_D$, but in the full temperature range, if $N_G\ll 1$
(otherwise, the integration over the frequency is essentially different from
the one-dimensional oscillator approximation because $|\nu(\omega)|$ depends
on the frequency).

We find the relation
\begin{equation}
 \label{T0}
      T_0=\frac{6}{\pi}\, H_\Lambda N_G=\frac{6}{\pi}\,\sqrt{\Omega_\Lambda}
      H_0N_G,
\end{equation}
wherefrom it follows that $T_0\ll H_0$ in consistency with the empirical data
setting $a_0 \approx H_0/2\pi$, so that
\begin{equation}
 \label{empir}
      N_G\approx\frac{1}{24\pi}\frac{1}{\sqrt{\Omega_\Lambda}}\sim 10^{-2}.
\end{equation}
One can see that the quantity $N_G$ is the new fundamental parameter in the
framework of thermodynamical and holographical gravity.

\section{Discussion and conclusion}

In our approach the thermodynamical ``bit'' on the holographic screen
occupies the Planckian area
$A_\mathrm{Pl}=\ell_\mathrm{Pl}\times\ell_\mathrm{Pl}$, defining the quantum
of area. Note that a binding energy of the screen ``bits'' is perhaps of the
order of Planck scale itself. To speculate on the notion of surface, one
should require that energy fluctuations in the transverse direction to the
screen should be much less than the Planck energy.

On the other hand, the ``bits'' on the screen experience the collective
transverse motion as a whole. Then, the quantity defined by
$$
      \Lambda_G=\frac{N_G}{\ell_\mathrm{Pl}},
$$
represents the density of vibrational modes per the Planckian length. By
dimension, it is the energy of transverse motion for the area quantum or
``bit''. Therefore, the surface structure is rigid or stable, if
$\Lambda_G\ll m_\mathrm{Pl}$.

Thus, we draw the conclusion that the stability of holographic screen as the
surface demands $N_G\ll 1$, while by the order of magnitude ``the transverse
energy of the Planckian quantum area'' is $\Lambda_G\approx
N_Gm_\mathrm{Pl}\sim 10^{16-17}$ GeV, i.e. it is close to the scale of the
grand unification theory (GUT).

For the Milgrom acceleration we obtain
\begin{equation}
 \label{ao2}
      a_0\approx 12\sqrt{\Omega_\Lambda}H_0\,\frac{\Lambda_G}{m_\mathrm{Pl}}.
\end{equation}
Thus, the basic parameter of the MOND involves the scale of vacuum energy
(the cosmological constant), Planck mass and the scale closed to the GUT
energy.

In this paper we have given the justification for the modification of the
equipartition rule for the ``bits'' on the holographic screen at low
temperatures due to the proper description for the low-frequency collective
transverse oscillations of screen at the presence of cosmological constant.
The requirement on the stability of holographic screen with respect to such
oscillations is reduced to the introduction of energy scale which should be
substantially less than the Planck energy. Then, the corrected
low-temperature distribution of thermal energy on the holographic screen is
reduced to the modified gravitational law like MOND at accelerations of free
fall less than the critical Milgrom acceleration. The Milgrom acceleration
has been empirically introduced in the MOND paradigm in order to describe the
flat rotation curves in spiral galaxies and to explain the Tully-Fisher law
connecting the visible mass of galaxy to the star velocity in the region
dominating by the dark matter halo. Phenomenologically, observational data
give the scale of transverse oscillating energy for the screen close to the
scale of grand unification theory.

Another treatment of Milgrom acceleration in terms of entropic force has been
recently presented in \cite{Ho:2010ca}.

\section*{Acknowledgments}

This work was partially supported by grants of Russian Foundations for Basic
Research 09-01-12123 and 10-02-00061, Special Federal Program ``Scientific
and academics personnel'' grant for the Scientific and Educational Center
2009-1.1-125-055-008, ant the work of T.S.A. was supported by the Russian
President grant MK-406.2010.2.

\appendix

\section{\label{app1}Transverse oscillations running along the screen}

\begin{figure}
\includegraphics[scale=0.9]
{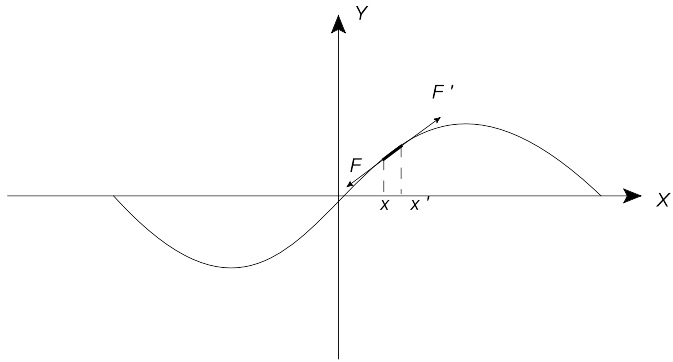} \caption{Lateral oscillations of the screen.}\label{ris}
\end{figure}

Let us derive the speed of propagation for the transverse oscillations on the
holographic screen with wave vector in the screen plane. The screen can be
considered as a membrane with the constant surface tension $\sigma=-p_A=Ts_A$	
and the surface density of energy $\rho_A=E/A=2Ts_A$ in the approximation of
high temperatures. Let us regard the sinusoidal wave running in the direction
of axis $X$ (Fig.\ref{ris}).

At the initial time the wave equation reads off
$$
y(x)=\mathcal{A} \sin \left(\frac{2\pi x}{\lambda}\right),
$$
where $\mathcal{A}$ is a small wave amplitude, $\lambda$ is a wave length.
Let us select the element of the wave between $x$ and $x '$ coordinates.
Forces $F$ and $F'$ are equal to each other by magnitude $\sigma L$, where
$L$ is a wave width in the direction perpendicular to the wave vector. But
the directions of those forces are tangent to the screen and, hence,
different because the points of action slightly differ. Denote  the angles
between axes $X$ and forces $F,\,F '$ by $\alpha$ and $\alpha '$,
respectively, and put $\Delta x=x'-x$. Then the mass of selected element
equals $m=\rho_A \Delta x L$, where we ignore the slope of the wave to axis
$X$, because the wave amplitude is small. Similarly, we neglect the net force
along axis $X$. The net force along axis $Y$ equals $\sigma L (\sin\alpha
-\sin\alpha ')=\sigma L \frac{\ud^2 y(x)}{\ud x^2}\Delta x= -\sigma L \Delta
x \left(\frac{2\pi}{\lambda}\right)^2 y$. Immediately, we obtain the equation
of oscillations
$$
\rho_A  \Delta x L \ddot{y}=-\sigma L \Delta x
\left(\frac{2\pi}{\lambda}\right)^2 y \qquad\Rightarrow\qquad
\ddot y+c_T^2 y=0.
$$
So, we straightforwardly get the square of the wave speed
$$c_T^2=\frac{\sigma}{\rho_A}=\frac12.$$

Notice, that considered oscillations occur at zero gradients of the energy
density and pressure, i.e. it is not a sound. Moreover, since the holographic
screen is the surface with a constant acceleration of free fall, these
oscillations take place in connection with proper oscillations of the {\it
matter} inside the screen.

\end{document}